\RequirePackage{fix-cm}
\documentclass[smallextended]{svjour3}       
\smartqed  
\usepackage{graphicx}
\usepackage{amsmath,epsfig,amssymb,graphicx,caption}
\usepackage{subcaption}
\usepackage{setspace}
\usepackage[colorlinks,citecolor=blue]{hyperref}
\usepackage{lineno}
\newcommand{\hSRone}{h_\mathrm{SR_1}}
\newcommand{\hSRtwo}{h_\mathrm{SR_2}}
\newcommand{\hRoneD}{h_\mathrm{R_1D}}
\newcommand{\hRtwoD}{h_\mathrm{R_2D}}
\newcommand{\hRR}{h_\mathrm{R_1R_2}}
\newcommand{\gammaSRavg}{\bar{\gamma}_\mathrm{SR}}
\newcommand{\gammaRDavg}{\bar{\gamma}_\mathrm{RD}}
\newcommand{\gammaRRavg}{\bar{\gamma}_\mathrm{IR}}
\newcommand{\gammaSR}{\gamma_\mathrm{SR}}
\newcommand{\gammaRD}{\gamma_\mathrm{RD}}
\newcommand{\gammaRR}{\gamma_\mathrm{IR}}

\begin{document}
\title{Power Allocation and Effective Capacity of AF Successive Relays}
\author{Mohammad Lari}
\institute{Electrical and Computer Engineering Faculty \at
              Semnan University, Semnan, Iran \\
              Tel.: +9823-33383947\\
              \email{m\_lari@semnan.ac.ir}}
\date{Received: date / Accepted: date}
\maketitle
\begin{abstract}
In the relay based telecommunications with $K$ relays between the source and destination, $K+1$ time or frequency slots are required for a single frame transmission. However, without the relays, only one time or frequency slot is used for a single frame transmission. Therefore, despite the benefits of relaying systems, this type of communications is not efficient from the spectral efficiency viewpoint. One solution to reduce this issue might be the full-duplex (FD) relays. An old technique which is reconsidered recently to improve the spectral efficiency of telecommunication systems. However, FD relays have a certain complexity, so, some similar techniques such as successive relays with nearly the same performance but less complexity is taken into account now. In successive relaying systems, two relays between the source and destination are employed which receive the transmitted frames from the source and relay it to the destination successively. This structure generally acts like an FD relays. In this paper, the effective capacity performance of an amplify and forward (AF) successive relaying systems with power allocation strategy at the relays are studied perfectly. However, while the inter-rely interference (IRI) between two successive relays has to be managed well, the power allocation and the effective capacity is derived under different assumptions about the IRI. In this way, we assume weak or strong, short or long-term constraints on the IRI. Then we extract the optimal transmitted power at the relay to maximize the effective capacity under these constraints.
\end{abstract}
\section{Introduction}\label{sec:1}
Full-Duplex (FD) communications is a promising technique which offers to double the spectral efficiency of radio links \cite{Tabataba_FD}. This impressive capability can response part of the explosive demands for high data-rate services. Due to this and despite multiple drawbacks of FD implementation, this technique has been considered as a candidate for next generation 5G wireless networks \cite{Zhang_FD}. In addition, FD can eliminate hidden terminal problem in ad-hoc networks, congestion, and large end-to-end delays \cite{Main_FD_1}. Therefore, FD communication systems have attracted many attentions in research area recently \cite{Main_FD_1}-\cite{Main_FD_2}.

Half-Duplex (HD) and FD are two ways for connecting terminals in a wireless network. Traditionally, HD terminals transmit or receive  either at different times or over different frequency bands. Therefore, for a simple connection, two time or frequency slots are consumed. On the other hand and when the terminals support FD connection, transmission and reception can be accomplished simultaneously in a same time slot and frequency band. Therefore, resource consumption is reduced to the half of traditional HD links \cite{Main_FD_1}-\cite{Main_FD_2}. Clearly, in a dense environment with a large number of users, employment of FD terminals can release a high amount of resources and may lead to the significant increase in data rate of active users. However, self-interference (SI) from a transmitter to its own receiver is the evident result of using FD terminals. Hence, management and reduction of SI are vital for the practical implementation of FD systems \cite{Main_FD_1}-\cite{Perez_Femtocell}. Fortunately, recent researches show efficient techniques for SI reduction which were not possible until a few years ago \cite{Duarte_FD}. So, new research like our present paper can be considered for the practical purpose completely.

Another challenging technology for next generation 5G wireless networks is relaying which can provide lower transmit powers, higher throughput, and extensive coverage \cite{AS_FD}. However, relaying suffers from low spectral efficiency. For example, consider one-way transmission from a source to a destination together with $K$ parallel relays. For complete transmission from the source and relays to the destination, $K+1$ different time  slots or frequency bands are required. This may be compared with direct transmission between the source and destination which uses only one time slot and frequency band. Here, FD devices (FD relays and FD destination) can be organized to reduce this weakness \cite{Ngoc_FD}. Therefore, FD relaying systems are of most interest among the other FD systems. 

To alleviate the complexity of FD relays, some quasi FD schemes like successive relaying, two-way relaying or buffer aided relaying are investigated \cite{Main_FD_1}, \cite{Main_SR_1}-\cite{Two_Way_Relaying}. These techniques try to mimic FD relaying via two HD relays. In successive relaying, two relays listen to the source and retransmit their received signal to the destination successively. In other words, in each time slot, one of the relays receives a new data frame from the source while the other relay forwards the previous data frame to the destination. Then, the role is swapped in the next time slot. In this way, a new data frame can be sent to the destination in each time slot as if FD relay was employed. In this situation, while transmission and reception of two relays occur in a same time slot and frequency band, the transmitted signal from one relay can disturb the received signal by the other relay simultaneously. Therefore, inter-relay interference (IRI) may degrade the whole performance. So, it has to be managed well in successive relaying \cite{Main_SR_2}-\cite{Main_SR_3}. 

Due to high performance and simplicity of successive relaying, this paper concentrates on the power allocation of this scheme. Power allocation is widely take into account in different relay structures such as amplify and forward (AF) \cite{Lim_AF}-\cite{Ngoc_AF} or decode and forward (DF) \cite{Kim_DF}-\cite{Cao_DF} relaying and under various conditions such as HD \cite{Farhadi_HD}-\cite{Ardebilipour_HD} or FD \cite{MILCOM_FD}-\cite{Riihonen_FD} transmission mode. In all these papers, the transmitted power at the source and/or relays is optimized to achieve certain target under power constraints. Several optimization targets have been adopted, such as outage probability minimization, capacity maximization, and signal to noise ratio (SNR) maximization. More precisely, successive relaying is also attracted attention and has been investigated from different aspects \cite{SR_New_1}-\cite{SR_New_6}. For example in \cite{SR_New_1}-\cite{SR_New_3}, the IRI suppression are studied and different schemes based on differential cancellation and network coding are proposed for two-path successive relays. In \cite{SR_New_4}-\cite{SR_New_5}, the authors have examined the throughput rate of a successive relay network which is suffered by the IRI. Then in \cite{SR_New_6}, a spectrum sharing scheme for overlaid wireless networks based on the DF successive relaying technique is proposed. However, power allocation and effective capacity in successive relaying in addition to IRI considerations is not investigated previously \footnote{For a complete review of successive relaying, please refer to \cite{Hanzo}}. Here, we study the power allocation of AF successive relays  for effective capacity maximization. The effective capacity is more general than the ergodic capacity which is defined as a maximum constant arrival rate with the delay quality-of-service (QoS) guarantee \cite{Lari1}-\cite{Lari4}. Then, the allocated power is derived under short-term and long-term IRI constraints and compared with different allocation techniques with different QoS requirements. IRI is not always harmful, and, strong IRI is more desirable in some states \cite{Main_SR_1}-\cite{Main_SR_3}. Therefore, in the following, we assume both weak and strong IRI in our optimization problem separately. 

This paper is organized as follows: First, the system model is introduced in Section \ref{sec:2}, where, successive relaying and interference management in this scheme are discussed. In addition, channel model and the effective capacity are explained here. In Section \ref{sec:3}, we plan our optimization problem to maximize the effective capacity under different power constraints and then, the optimal allocated power is obtained in this section. Finally, the simulation results are presented in Section \ref{sec:4}, and Section \ref{sec:5} concludes the paper.
\section{System Model}\label{sec:2}
\subsection{Successive Relaying}\label{subsec:21}
\begin{figure}[h]
\begin{center}
\includegraphics[draft=false,scale=1]{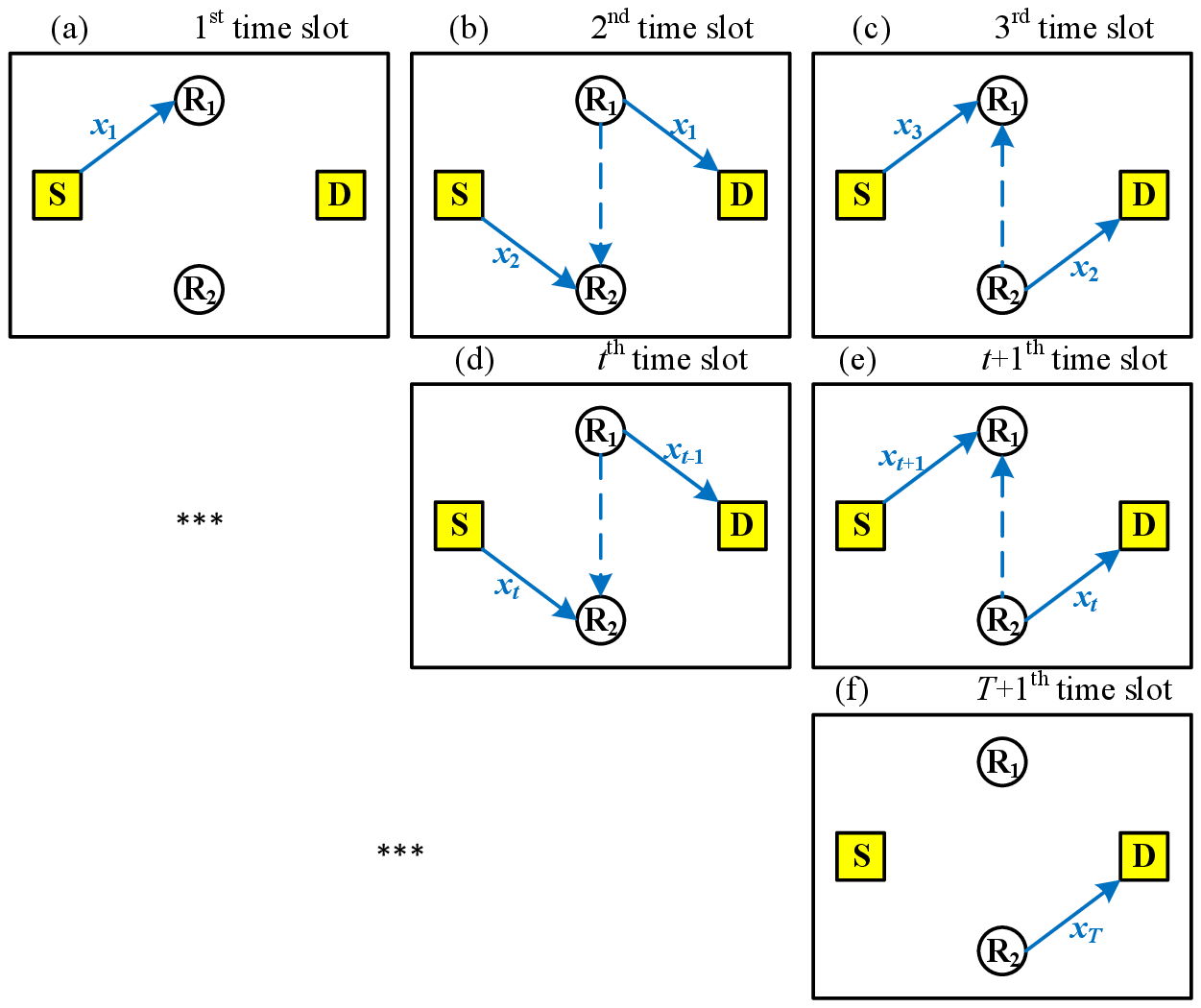}
\end{center}
\caption{Successive relaying in different time slots. Solid lines depicts the main paths and dashed lines show the interferer paths.}
\label{fig:successive_relaying}
\end{figure}

Successive relaying behavior is shown in Fig. \ref{fig:successive_relaying}, where two AF relay nodes $\mathrm{R_1}$ and $\mathrm{R_2}$, cooperate the source node $\mathrm{S}$, amplify its received signal and retransmit it to the destination $\mathrm{D}$ successively. The direct source to destination link is also ignored. For more details, we assume each data frame has $T$ symbols $\{x_1,x_2,...,x_T\}$, which has to be transmitted from the source to the destination. The specific steps for each time slot are described as follows \cite{Main_SR_2}: 
\begin{itemize}
\item{In the $1^{\mathrm{st}}$ time slot, $\mathrm{S}$ transmits $x_1$, $\mathrm{R_1}$ listens to $x_1$ from $\mathrm{S}$ and $\mathrm{R_2}$ is silent (see Fig. \ref{fig:successive_relaying}(a)).}
\item{In the $2^{\mathrm{nd}}$ time slot, $\mathrm{S}$ transmits $x_2$, $\mathrm{R_2}$ listens  to $x_2$ from $\mathrm{S}$ while being interfered by $x_1$ from $\mathrm{R_1}$ and $\mathrm{R_1}$ amplifies and forwards $x_1$ to $\mathrm{D}$ (see Fig. \ref{fig:successive_relaying}(b)).}
\item{In the $3^{\mathrm{rd}}$ time slot, $\mathrm{S}$ transmits $x_3$, $\mathrm{R_1}$ listens  to $x_3$ from $\mathrm{S}$ while being interfered by $x_2$ from $\mathrm{R_2}$ and $\mathrm{R_2}$ amplifies and forwards $x_2$ to $\mathrm{D}$ (see Fig. \ref{fig:successive_relaying}(c)).}
\item{In the $t^{\mathrm{th}}$ time slot ($t$ is even), $\mathrm{S}$ transmits $x_t$, $\mathrm{R_2}$ listens  to $x_t$ from $\mathrm{S}$ while being interfered by $x_{t-1}$ from $\mathrm{R_1}$ and $\mathrm{R_1}$ amplifies and forwards $x_{t-1}$ to $\mathrm{D}$ (see Fig. \ref{fig:successive_relaying}(d)).}
\item{In the $t+1^{\mathrm{th}}$ time slot ($t+1$ is odd), $\mathrm{S}$ transmits $x_{t+1}$, $\mathrm{R_1}$ listens  to $x_{t+1}$ from $\mathrm{S}$ while being interfered by $x_t$ from $\mathrm{R_2}$ and $\mathrm{R_2}$ amplifies and forwards $x_t$ to $\mathrm{D}$ (see Fig. \ref{fig:successive_relaying}(e)).}
\item{Finally, in the ${T+1}^{\mathrm{th}}$ time slot ($T+1$ is odd), $\mathrm{R_2}$ amplifies and forwards $x_{T}$ to $\mathrm{D}$ (see Fig. \ref{fig:successive_relaying}(f)).}
\end{itemize}
Now, we can observe that $T$ symbols are transmitted from the source and received completely in $T+1$ time slots by the destination. So, the multiplexing ratio becomes $T/(T+1)$ which approaches $1$ for a large number of symbols $T$. This leads to the high spectral efficiency as can be obtained in FD relaying systems. Except for the first and last time slots, relays provide interference for each other. Note that, for achieving high performance of relaying systems, this IRI has to be managed carefully.
\subsection{Interference Management}\label{subsec:22}
For proper relaying system performance, management of IRI between two relays is essential. Hence, applying interference cancellation techniques such as interference alignment (IA) \cite{Zhao_1}-\cite{Zhao_2} or successive interference cancellation (SIC) \cite{Main_SR_2} is strongly recommended. However for the sake of simplicity, we assume that the relays using SIC for IRI suppression and we explain the interference cancellation as follows: (a) When the IRI is weak, the  relay does not attempt to  remove the interference. So, the power of interference is added to the power of noise and the overall SNR is reduced slightly. (b) On the other hand, when the IRI is strong, the relay first detects the interferer signal. Then, subtracts the estimated interference from the received signal. Now, detection of the main signal without interference is possible. In this case, there will be no SNR reduction due to interference. Hence, the strong IRI is not damaging in this case. 

Most of the time, interference is detrimental and interferer signal and symbols are discarded. However, this subject is not always the best solution. In our successive relaying scenario, the interference signal can be detected and used for further improvements. For example, according to description of \ref{subsec:21}, $\mathrm{R_1}$ can extract just $\{x_1,x_3,...\}$ and $\mathrm{R_2}$ can extract just $\{x_2,x_4,...\}$ when they transmitted from the source. Hence,  these two relays do not have an access to all symbols completely. But, by utilization of the IRI, $\mathrm{R_1}$ and $\mathrm{R_2}$ can  obtain $\{x_2,x_4,...\}$ and $\{x_1,x_3,...\}$ respectively. Well, using the complete set of symbols in the relays, may lead to higher performance \footnote{The complete set of symbols can be used for channel estimation, higher order of diversity or ..., however, the details and method of using IRI is not our concern here in this paper.}. Therefore, the strong IRI can be helpful in this scheme. For this reason, the strong IRI will be considered as one of the constraints in the optimization problem of section \ref{sec:3}.
\subsection{Channel Model}\label{subsec:23}
Here, we assume uncorrelated quasi-static flat fading channel, where, $\hSRone$ and $\hSRtwo$ denote $\mathrm{S}-\mathrm{R_1}$ and $\mathrm{S}-\mathrm{R_2}$ channel coefficients and $\hRoneD$ and $\hRtwoD$ show $\mathrm{R_1}-\mathrm{D}$ and $\mathrm{R_2}-\mathrm{D}$ channel coefficients and $\hRR$ determines $\mathrm{R_1}-\mathrm{R_2}$ channel coefficient respectively. The distances of $\mathrm{S}-\mathrm{R_1}$ and $\mathrm{S}-\mathrm{R_2}$ as well as  $\mathrm{R_1}-\mathrm{D}$ and $\mathrm{R_2}-\mathrm{D}$ are nearly the same. Also, the average transmitted power by the source and each relay are assumed $P_\mathrm{S}$ and $P_\mathrm{R}$ and the power of additive white Gaussian noise (AWGN) is the same at the relays and destination. Therefore, the average SNR at the $\mathrm{S}-\mathrm{R_1}$ or $\mathrm{S}-\mathrm{R_2}$ link is $\gammaSRavg$ and at the $\mathrm{R_1}-\mathrm{D}$ or $\mathrm{R_2}-\mathrm{D}$ link is expressed as $\gammaRDavg$ and at the $\mathrm{R_1}-\mathrm{R_2}$ link is written as $\gammaRRavg$. Now, assuming IRI cancellation with the SIC technique, the overall instantaneous SNR at the destination is written as \cite{Collings}
\begin{equation}\label{eq:1}
\gamma_{\mathrm{eq}}=\frac{\gammaSR\gammaRD}{\gammaSR+\gammaRD+1}
\end{equation}
where at the even time slots, $\gammaSR=\gammaSRavg|\hSRone|^2$ and $\gammaRD=\gammaRDavg|\hRoneD|^2$ and at the odd time slots $\gammaSR=\gammaSRavg|\hSRtwo|^2$ and $\gammaRD=\gammaRD|\hRtwoD|^2$ respectively. In this way, the SNR at the link between the relays can be expressed as $\gammaRR=\gammaRRavg|\hRR|^2$. In the Rayleigh fading channels, where $\gammaSRavg=\gammaRDavg=\bar{\gamma}$, the probability density function (PDF) and the cumulative distribution function (CDF) of the SNR are expressed as \cite{Collings}
\begin{equation}\label{eq:2}
f_{\gamma_{\mathrm{eq}}}(x)=e^{-2x/\bar{\gamma}}\frac{4x}{\bar{\gamma}^2}\left[K_1\left(\frac{2x}{\bar{\gamma}}\right)+K_0\left(\frac{2x}{\bar{\gamma}}\right)\right]
\end{equation}
\begin{equation}\label{eq:3}
F_{\gamma_\mathrm{eq}}(x)=1-e^{-2x/\bar{\gamma}}\frac{2x}{\bar{\gamma}}K_1\left(\frac{2x}{\bar{\gamma}}\right)
\end{equation}
where $K_1(.)$ and $K_0(.)$ are the first and zero-order modified Bessel function of the first kind respectively \cite[eq. 8.407]{Ryzhik}.
\subsection{Effective Capacity}\label{subsec:24}
The effective capacity is a similar concept to the capacity which  investigates the effect of transmitter and receiver buffer (queue) delay on the capacity \cite{Negi}-\cite{Soret}. In this regard, the effective capacity is defined as the maximum arrival rate that a wireless channel can support, in order to guarantee the QoS requirements such as the statistical delay constraint \cite{Negi}-\cite{Soret}. Consequently, the effective capacity is more realistic and can analyse the performance of the wireless channels against delay sensitive traffics more precisely.

For a dynamic queuing system with stationary and ergodic arrival and service processes, under sufficient conditions, the queue length process $Q(t)$ converges in distribution to a random variable $Q(\infty)$ such that \cite{Negi}
\begin{equation}\label{eq:4}
\textrm{Pr}\{Q(\infty)>q\}\approx \varepsilon e^{-\theta q}
\end{equation}
for a large $q$, where $\varepsilon$ represents the non-empty buffer probability, $q$ determines a certain threshold and $\theta$ indicates the amount of required QoS. In addition, for the delay of a packet in the buffer as the main QoS metric, we have a similar probability function as \cite{Negi}
\begin{equation}\label{eq:5}
\textrm{Pr}\{D>d\}\approx\varepsilon e^{-\theta \delta d},
\end{equation}
where $D$ indicates the tolerated delay, $d$ is a delay-bound, and $\delta$ is jointly determined by both arrival and service processes. The statistical delay constraint in \eqref{eq:5} represents the QoS which has to be guaranteed for the delay sensitive traffic sources. It is apparent that the QoS exponent $\theta$ has an important role here. Larger $\theta$ corresponds to more strict QoS constraint, while smaller $\theta$ implies looser QoS requirements.

Effective capacity provides the maximum constant arrival rate that can be supported by the time-varying wireless channel under the statistical delay constraint \eqref{eq:5}. Since the average arrival rate is equal to the average departure rate when the buffer is in a steady-state \cite{Negi}, the effective capacity can be viewed as the maximum throughput in the presence of such a constraint. With all these explanations, the effective capacity is defined as \cite{Negi}-\cite{Soret} 
\begin{equation}\label{eq:6}
E_C(\theta)=-\frac{1}{\theta}\ln\left(\textsf{E}\left\{e^{-\theta R}\right\}\right)
\end{equation}
where $R$ is the time-varying rate of the channel and $\textsf{E}\{.\}$ denotes the statistical expectation. For a specific application with a given statistical delay requirement, the QoS exponent $\theta$ can be determined from \eqref{eq:5}. Then, the maximum constant arrival rate of the sources that a wireless channel can support in order to guarantee the given QoS is concluded from \eqref{eq:6}. The power allocation scheme and the effective capacity of AF successive relaying are discussed in the next section.
\section{Power Allocation Strategy}\label{sec:3}
In order to maximize the effective capacity, the allocation power scheme at the relay is extracted here. As we explained before, high performance of successive relaying is obtained when the IRI is considered and managed well. So, we will assume weak or strong and short-term or long-term IRI constraints in our optimization problem. Note that, weak constraint means the IRI is lower than the specific threshold $q_0$ and subsequently, strong constraint means the IRI is higher than the specific threshold $q_0$. Besides, short-term constraint means the instantaneous IRI and long-term constraint means the average IRI respectively. 

With power allocation strategy, we assume that the relays can adapt their transmitted power to maximize the effective capacity. In this regard, the relays transmit $\mu_0P_\mathrm{R}$ instead of $P_\mathrm{R}$, where $\mu_0\geq0$ depicts the time-varying allocated power coefficient and we have $\textsf{E}\{\mu_0\}=1$ for constant average transmitted power. Now, with this policy, the instantaneous SNR at the destination is changed to 
\begin{equation}\label{eq:7}
\tilde{\gamma}_{\mathrm{eq}}=\frac{\gammaSR\mu_0\gammaRD}{\gammaSR+\mu_0\gammaRD+1}
\end{equation}
and the instantaneous capacity is written as
\begin{equation}\label{eq:8}
R=B\left(\frac{T}{T+1}\right)T_0\log_2(1+\tilde{\gamma}_{\mathrm{eq}})
\end{equation}
where $B$ specify the total required bandwidth and $T_0$ denotes one symbol duration or one transmission time slot in Fig. \ref{fig:successive_relaying}. Since $T$ symbols are transmitted from the source to the destination in $T+1$ time slots, the instantaneous capacity is written as \eqref{eq:8}. By substituting \eqref{eq:8} into \eqref{eq:6}, the effective capacity of AF successive relays is obtained as
\begin{eqnarray}\label{eq:9}
E_C(\theta)&=&-\frac{1}{\theta}\ln\left(\textsf{E}\left\{e^{-\theta B\left(\frac{T}{T+1}\right)T_0\log_2(1+\tilde{\gamma}_{\mathrm{eq}})}\right\}\right)\nonumber\\
&=&-\frac{1}{\theta}\ln\left(\textsf{E}\left\{e^{-\tilde{\theta}\ln(1+\tilde{\gamma}_{\mathrm{eq}})}\right\}\right)\nonumber\\
&=&-\frac{1}{\theta}\ln\left(\textsf{E}\left\{(1+\tilde{\gamma}_{\mathrm{eq}})^{-\tilde{\theta}}\right\}\right)
\end{eqnarray}
where $\tilde{\theta}=\theta B\left(\frac{T}{T+1}\right)T_0/\ln 2$. As we explained before, the IRI has a significant influence in the performance of successive relays and this interference has to be managed properly. When the weak IRI is desirable, we can adjust the transmitted power of the relays to obtain the short-term or long-term interference below some specific threshold $q_0$. Consequently, we use $\mu_0\gamma_{\mathrm{IR}}\leq q_0$ as the short-term constraint or $\textsf{E}\{\mu_0\gamma_{\mathrm{IR}}\}\leq q_0$ for the long-term constraint as well. As the same way, when we need the strong IRI, we can adapt the transmitted power from the relays to obtain the short-term or long-term interference above the threshold $q_0$. Therefore, we would have $\mu_0\gamma_{\mathrm{IR}}\geq q_0$ as the short-term constraint or $\textsf{E}\{\mu_0\gamma_{\mathrm{IR}}\}\geq q_0$ for the long-term constraint, respectively. Now, we can summarize our optimization problem as 
\begin{eqnarray}\label{eq:10}
\mu_0^{\mathrm{opt}}&=&\arg\mathop{\max}_{\mu_0}-\frac{1}{\theta}\ln\left(\textsf{E}\left\{(1+\tilde{\gamma}_{\mathrm{eq}})^{-\tilde{\theta}}\right\}\right)\nonumber\\
&=&\arg\mathop{\max}_{\mu_0}-\frac{1}{\theta}\ln\left(\textsf{E}\left\{\left(1+\frac{\gammaSR\mu_0\gammaRD}{\gammaSR+\mu_0\gammaRD+1}\right)^{-\tilde{\theta}}\right\}\right)
\end{eqnarray}
\begin{equation}\label{eq:11}
\mathrm{s.t.}\quad\quad\mu_0\geq0
\end{equation}
\begin{equation}\label{eq:12}
\quad\quad\quad\textsf{E}\{\mu_0\}=1
\end{equation}
\begin{subequations}\label{eq:13}
\begin{align}
\label{eq:13a}
\quad\quad\quad\mu_0\gamma_{\mathrm{IR}}\leq q_0\\                                                                                                                                                                                                                                                                                                                                                                                                                                                                                                                                                                                                                                                                                                                                                                                                                                                                                                                                                                                                                                                                                                                                                                                                                                                                                                                                                                                                                                                                                                                                                                                                                                                                                                                                                                                                                                                                                                                                                                                                                                                                                                                                                                                                                                                                                                                                                                                                                                                                                                                                                                                                                                                                                                                                                                                                                                                                                                                                                                                                                                                                                                                                                                                                                                                                                                                                                                                                                                                                                                                                                                                                                                                                                                                                                                  
\label{eq:13b}
\quad\quad\quad\textsf{E}\{\mu_0\gamma_{\mathrm{IR}}\}\leq q_0\\
\label{eq:13c}
\quad\quad\quad\mu_0\gamma_{\mathrm{IR}}\geq q_0\\
\label{eq:13d}
\quad\quad\quad\textsf{E}\{\mu_0\gamma_{\mathrm{IR}}\}\geq q_0
\end{align}
\end{subequations}
where \eqref{eq:11} and \eqref{eq:12} show the positive and constant average transmitted power constraints and \eqref{eq:13} depicts the short-term and long-term IRI constraints respectively (\eqref{eq:13a} shows the weak and short-term constraint, \eqref{eq:13b} shows the weak and long-term constraint, \eqref{eq:13c} shows the strong and short-term constraint and finally \eqref{eq:13d} shows the strong and long-term constraint). Using the Lagrangian optimization method \cite{Opt}, the problem can be solved and $\mu_0^{\mathrm{opt}}$ is derived. However, applying the Lagrange method in the optimization problem of \eqref{eq:10}, a high non-linear equation of $\mu_0^{\mathrm{opt}}$ is obtained and extraction of a closed-form solution for $\mu_0^{\mathrm{opt}}$ is not possible \cite{Lari1}. Therefore, we continue with a novel approximation and replace $\mu_0$ at the denominator of \eqref{eq:10} by its average $\textsf{E}\{\mu_0\}=1$. Then, we can rewrite \eqref{eq:10} as
\begin{equation}\label{eq:14}
\mu_0^{\mathrm{opt}}=\arg\mathop{\max}_{\mu_0}-\frac{1}{\theta}\ln\left(\textsf{E}\left\{\left(1+\frac{\gammaSR\mu_0\gammaRD}{\gammaSR+1\times\gammaRD+1}\right)^{-\tilde{\theta}}\right\}\right)
\end{equation}
and simplify it to
\begin{equation}\label{eq:15}
\mu_0^{\mathrm{opt}}=\arg\mathop{\max}_{\mu_0}-\frac{1}{\theta}\ln\left(\textsf{E}\left\{\left(1+\mu_0 \gamma_{\mathrm{eq}}\right)^{-\tilde{\theta}}\right\}\right)
\end{equation}
where $\gamma_{\mathrm{eq}}$ is introduced in \eqref{eq:1}. The constraints of the optimization remain unchanged. Now, we can obtain a closed-form solution for the allocated power in the following subsections.
\subsection{Long-term IRI constraints}\label{subsec:31}
Now, we can use \eqref{eq:11}, \eqref{eq:12} and \eqref{eq:13b} for the weak and long-term IRI constraints. Here, the optimal power allocation is determined as
\begin{equation}\label{eq:16}
\mu_0^{\mathrm{opt}}=
\begin{cases}
0&,\gamma_{\mathrm{eq}}<\gamma_{\textrm{T}}\\
\left(\frac{1}{\gamma_{\textrm{T}}}\right)^\frac{1}{\tilde{\theta}+1}\left(\frac{1}{\gamma_{\mathrm{eq}}}\right)^\frac{\tilde{\theta}}{\tilde{\theta}+1}-\frac{1}{\gamma_{\mathrm{eq}}}&,\gamma_{\mathrm{eq}}\geq \gamma_{\textrm{T}}
\end{cases}
\end{equation}
where $\gamma_{\textrm{T}}$ is a cut-off SNR threshold which is calculated from
\begin{equation}\label{eq:17}
\textsf{E}\left\{\mu_0^{\mathrm{opt}}\right\}=
\begin{cases}
\frac{q_0}{\bar{\gamma}}&,q_0<\bar{\gamma}\\1&,q_0\geq\bar{\gamma}
\end{cases}.
\end{equation}
Against, for the strong and long-term IRI constraints, we use \eqref{eq:11}, \eqref{eq:12} and \eqref{eq:13d} and obtain
\begin{equation}\label{eq:18}
\mu_0^{\mathrm{opt}}=
\begin{cases}
0&,\gamma_{\mathrm{eq}}<\gamma_{\textrm{T}}\\
\left(\frac{1}{\gamma_{\textrm{T}}}\right)^\frac{1}{\tilde{\theta}+1}\left(\frac{1}{\gamma_{\mathrm{eq}}}\right)^\frac{\tilde{\theta}}{\tilde{\theta}+1}-\frac{1}{\gamma_{\mathrm{eq}}}&,\gamma_{\mathrm{eq}}\geq \gamma_{\textrm{T}}
\end{cases}
\end{equation}
for the $q_0\leq \bar{\gamma}$. Note that, when $q_0 >\bar{\gamma}$, the power allocation with all constraints \eqref{eq:11}, \eqref{eq:12} and \eqref{eq:13d} is impossible. The cut-off SNR threshold $\gamma_{\textrm{T}}$ is calculatde from $\textsf{E}\{\mu_0\}=1$ (the proof of \eqref{eq:16} has been derived in Appendix \ref{appx:1} and the proof of \eqref{eq:18} is similar to Appendix \ref{appx:1}).

Finally, we can replace \eqref{eq:16} or \eqref{eq:18} into \eqref{eq:15} to obtain the effective capacity of AF successive relays with the weak or strong long-term IRI constraints as follow (for the proof, please see Appendix \ref{appx:2})
\begin{eqnarray}\label{eq:19}
E_C(\theta)&=&-\frac{1}{\theta}\ln\left\{\frac{\sqrt{\pi}}{2\Gamma(5/2)}\left[F(3,3/2,5/2;0)+\frac{1}{2}F(2,1/2,5/2;0)\right]\right.\nonumber\\
&-&\frac{\sqrt{\pi}\gamma_{\textrm{T}}}{\bar{\gamma}}\left[G_{23}^{30}\left(\frac{4\gamma_{\textrm{T}}}{\bar{\gamma}}\left|\begin{array}{l}0,3/2\\-1,2,0\end{array} \right.\right) + G_{23}^{30}\left(\frac{4\gamma_{\textrm{T}}}{\bar{\gamma}}\left|\begin{array}{l}0,3/2\\-1,1,1\end{array} \right.\right)\right]\nonumber\\
&+&\left.\frac{\sqrt{\pi}}{4}\left(\frac{4\gamma_{\textrm{T}}}{\bar{\gamma}}\right)^{\frac{1+2\tilde{\theta}}{1+\tilde{\theta}}}\left[G_{23}^{30}\left(\frac{4\gamma_{\textrm{T}}}{\bar{\gamma}}\left|\begin{array}{l}0,\frac{1}{2}+\frac{1}{1+\tilde{\theta}}\\-1,1+\frac{1}{1+\tilde{\theta}},-1+\frac{1}{1+\tilde{\theta}}\end{array} \right.\right)\right.\right.\nonumber\\
&+&\left.\left.G_{23}^{30}\left(\frac{4\gamma_{\textrm{T}}}{\bar{\gamma}}\left|\begin{array}{l}0,\frac{1}{2}+\frac{1}{1+\tilde{\theta}}\\-1,\frac{1}{1+\tilde{\theta}},\frac{1}{1+\tilde{\theta}}\end{array} \right.\right)\right]\right\}
\end{eqnarray}
where $F(.,.;.;.)$ represents Gauss hypergeometric function \cite[eq. 9.10]{Ryzhik}, $\Gamma(.)$ denotes the Gamma function and $G_{p,q}^{m,n}(.)$ is the Meijer's G function defined in \cite[eq. 9.301]{Ryzhik} which is easily evaluated using the popular numerical softwares.
\subsection{Short-term IRI constraints}\label{subsec:32}
When the weak and short-term IRI is required, the constraints of \eqref{eq:11}, \eqref{eq:12} and \eqref{eq:13a} can be used and the optimal allocated power is obtained as
\begin{equation}\label{eq:20}
\mu_0^{\mathrm{opt}}=
\begin{cases}
0&,\gamma_{\mathrm{eq}}<\gamma_{\textrm{T}}\\
\left(\frac{1}{\gamma_{\textrm{T}}}\right)^\frac{1}{\tilde{\theta}+1}\left(\frac{1}{\gamma_{\mathrm{eq}}}\right)^\frac{\tilde{\theta}}{\tilde{\theta}+1}-\frac{1}{\gamma_{\mathrm{eq}}}&,\gamma_{\mathrm{eq}}\geq \gamma_{\textrm{T}},\gammaRR<\gammaRR^*\\
\frac{q_0}{\gammaRR}&,\gamma_{\mathrm{eq}}\geq \gamma_{\textrm{T}},\gammaRR\geq\gammaRR^*
\end{cases}.
\end{equation}
On the other hand, with the strong and short-term IRI assumption, the constraints of \eqref{eq:11}, \eqref{eq:12} and \eqref{eq:13c} can be used to calculate the optimal allocated power as
\begin{equation}\label{eq:21}
\mu_0^{\mathrm{opt}}=
\begin{cases}
0&,\gamma_{\mathrm{eq}}<\gamma_{\textrm{T}}\\
\frac{q_0}{\gammaRR}&,\gamma_{\mathrm{eq}}\geq \gamma_{\textrm{T}},\gammaRR<\gammaRR^*\\
\left(\frac{1}{\gamma_{\textrm{T}}}\right)^\frac{1}{\tilde{\theta}+1}\left(\frac{1}{\gamma_{\mathrm{eq}}}\right)^\frac{\tilde{\theta}}{\tilde{\theta}+1}-\frac{1}{\gamma_{\mathrm{eq}}}&,\gamma_{\mathrm{eq}}\geq \gamma_{\textrm{T}},\gammaRR\geq\gammaRR^*
\end{cases}.
\end{equation}
Here, $\gamma_{\textrm{T}}$ is a cut-off SNR threshold which is determined from the average transmitted power, $\textsf{E}\{\mu_0\}=1$, and $\gammaRR^*$ is \footnote{The proof of \eqref{eq:21} and \eqref{eq:22} is very similar to Appendix \ref{appx:1}.}
\begin{equation}\label{eq:22}
\gammaRR^*=\frac{q_0}{\left(\frac{1}{\gamma_{\textrm{T}}}\right)^\frac{1}{\tilde{\theta}+1}\left(\frac{1}{\gamma_{\mathrm{eq}}}\right)^\frac{\tilde{\theta}}{\tilde{\theta}+1}-\frac{1}{\gamma_{\mathrm{eq}}}}.
\end{equation}
Finally, \eqref{eq:20} or \eqref{eq:21} can be replaced into \eqref{eq:15} and the effective capacity of AF successive relays with the weak or strong short-term IRI constraints is obtained. Here, finding a closed-form solution for the effective capacity is not straightforward. 
\section{Simulation Results}\label{sec:4}
For the simulation we assume $B=100\mathrm{KHz}$, $T_0=1\textrm{msec}$ and $\frac{T}{T+1}\approx 1$ and for more simple drawing, the normalized effective capacity $\overline{E_C(\theta)}=E_C(\theta)/(BT_0)$ is plotted in the following figures. 

\begin{figure}[h]
\begin{center}
\includegraphics[draft=false,width=\linewidth]{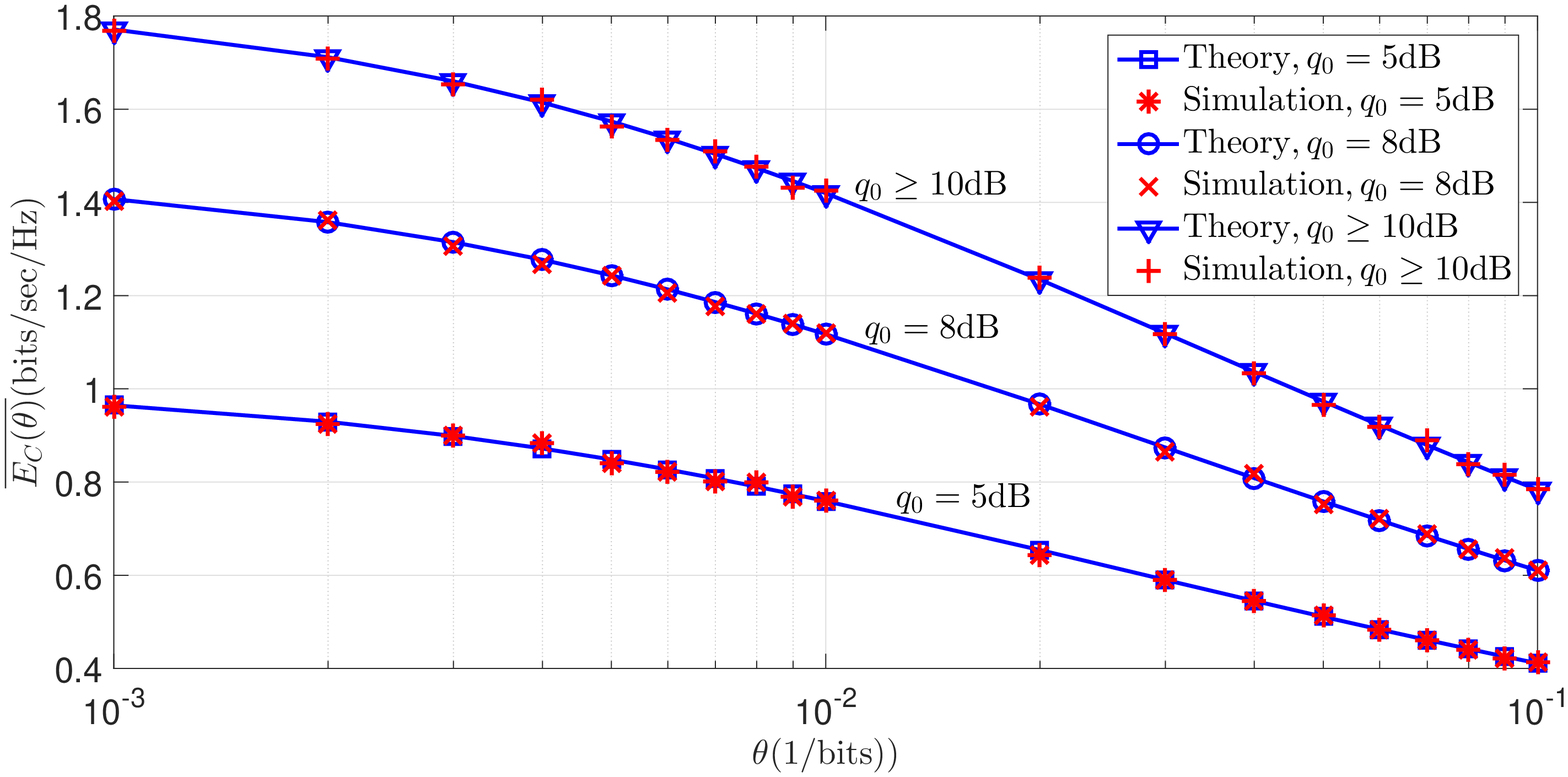}
\end{center}
\caption{Effective capacity in successive AF relays with the weak and long-term IRI constraints. We assume $\bar{\gamma}=10\textrm{dB}$ with $q_0=5\textrm{dB}$, $q_0=8\textrm{dB}$ and  $q_0\geq10\textrm{dB}$ here.}
\label{fig:2}
\end{figure}

\begin{figure}[h]
\begin{center}
\includegraphics[draft=false,width=\linewidth]{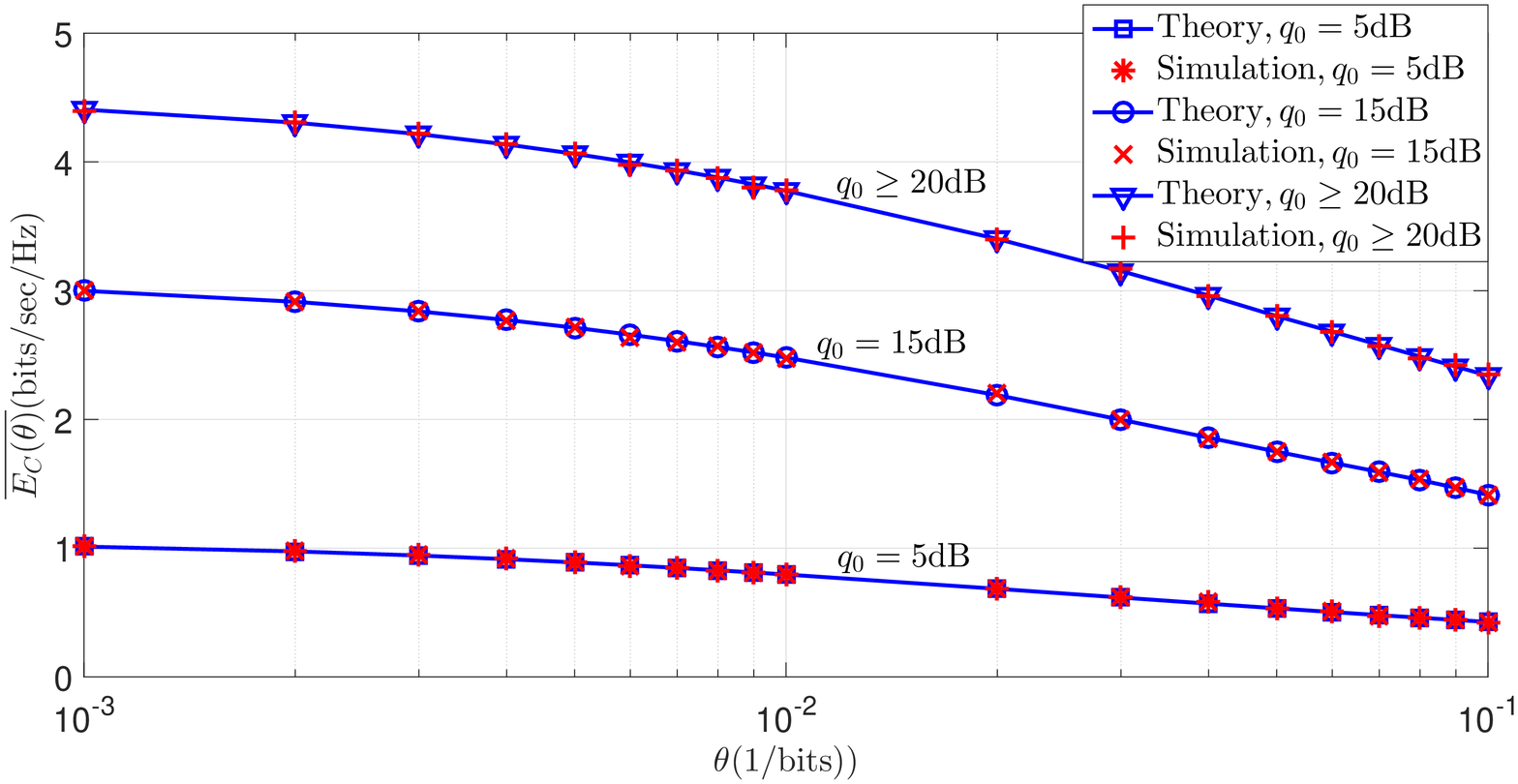}
\end{center}
\caption{Effective capacity in successive AF relays with the weak and long-term IRI constraints. We assume $\bar{\gamma}=20\textrm{dB}$ with $q_0=5\textrm{dB}$, $q_0=15\textrm{dB}$ and  $q_0\geq20\textrm{dB}$ here.}
\label{fig:3}
\end{figure}

The effective capacity of AF successive relays with the weak and long-term IRI constraints (see eq. \eqref{eq:13b}) is plotted in Fig. \ref{fig:2} when $\bar{\gamma}=10\textrm{dB}$ with $q_0=5\textrm{dB}$, $q_0=8\textrm{dB}$ and  $q_0\geq10\textrm{dB}$ and also in Fig. \ref{fig:3} when we have $\bar{\gamma}=20\textrm{dB}$ and $q_0=5\textrm{dB}$, $q_0=15\textrm{dB}$ and $q_0\geq20\textrm{dB}$ respectively. We can see the tight agreement between theory and simulation results in these figures which verifies the derived closed-form solution of the effective capacity in section \ref{subsec:31}. In addition, as we expected, the effective capacity increases when the threshold $q_0$ increases. Note that, when $q_0$ increases, the amount of tolerable interference between relays increases as well and therefore, the effective capacity of AF successive relay is increased.

\begin{figure}[h]
\begin{center}
\includegraphics[draft=false,width=\linewidth]{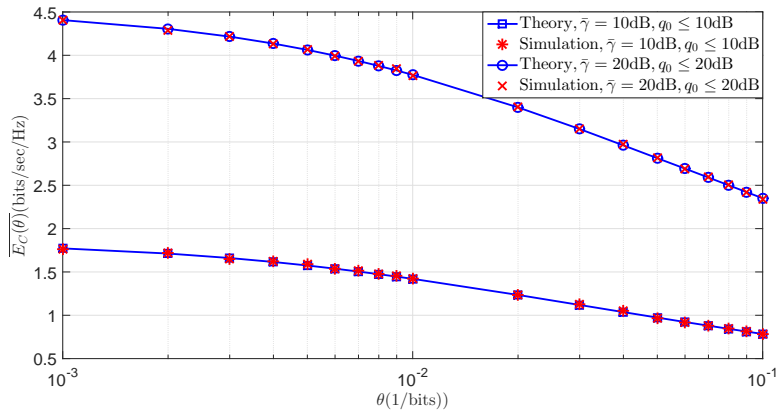}
\end{center}
\caption{Effective capacity in successive AF relays with the strong and long-term IRI constraints. We assume $\bar{\gamma}=10\textrm{dB}$ with $q_0\leq 10\textrm{dB}$ and  $\bar{\gamma}=20\textrm{dB}$ with $q_0\leq 20\textrm{dB}$ here.}
\label{fig:4}
\end{figure}

In a similar way, the effective capacity of AF successive relays with the strong and long-term IRI constraints (see eq. \eqref{eq:13d}) is plotted in Fig. \ref{fig:4}. Note that, the power allocation is not possible when $q_0>\bar{\gamma}$, because the strong constrain is not realized in this situation. Full agreement between theory and simulation is also clear in Fig. \ref{fig:4}.

\begin{figure}[h]
\begin{center}
\includegraphics[draft=false,width=\linewidth]{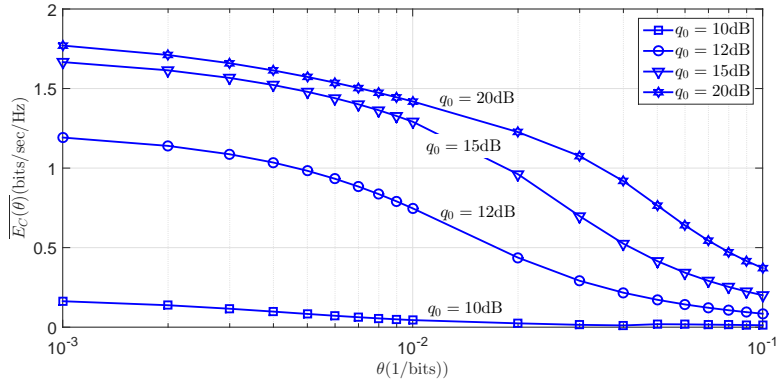}
\end{center}
\caption{Effective capacity in successive AF relays with the weak and short-term IRI constraints with $\bar{\gamma}=10\textrm{dB}$ and $q_0=10\textrm{dB}$, $q_0=12\textrm{dB}$, $q_0=15\textrm{dB}$ and $q_0=20\textrm{dB}$.}
\label{fig:5}
\end{figure}

\begin{figure}[h]
\begin{center}
\includegraphics[draft=false,width=\linewidth]{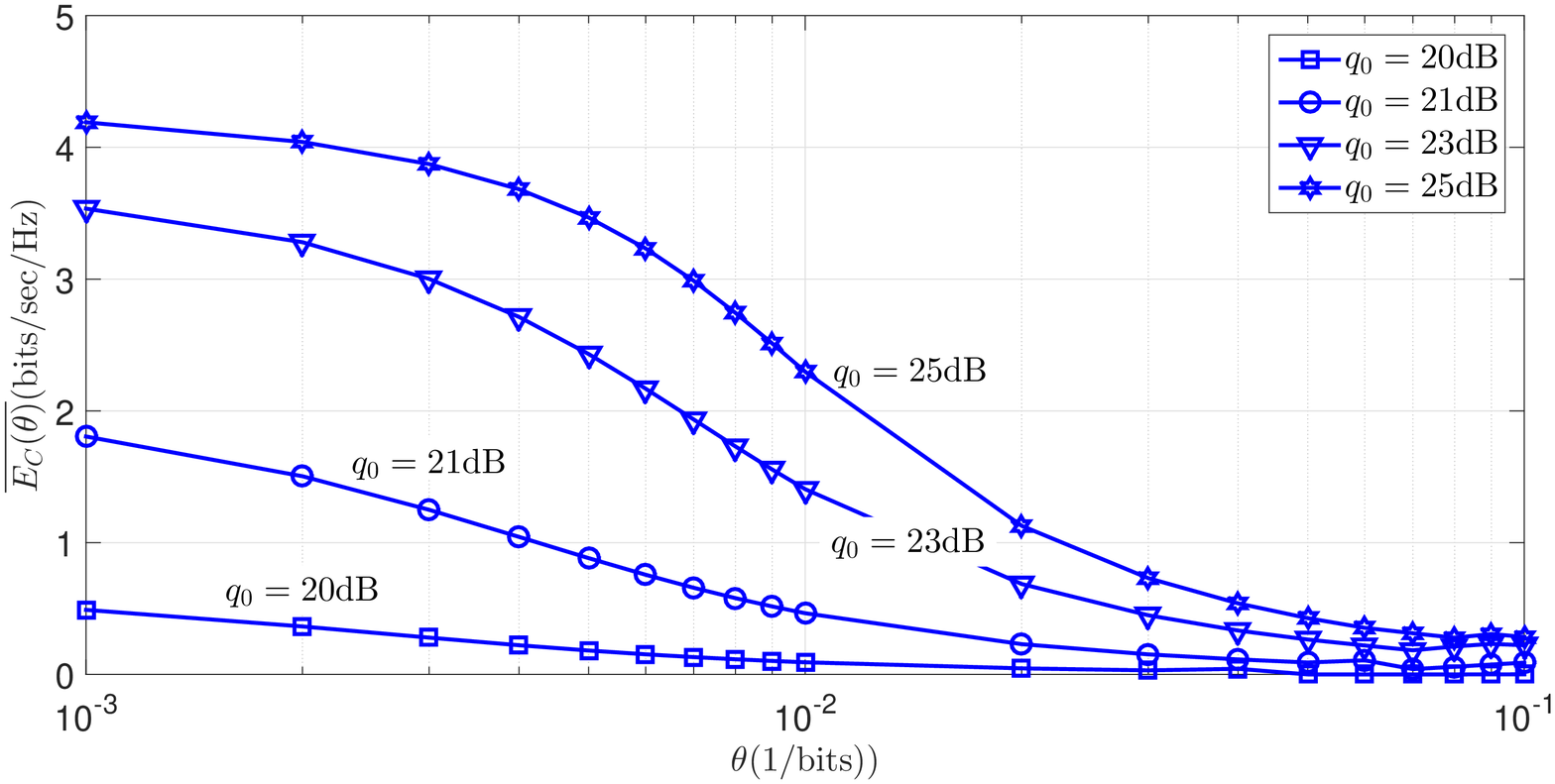}
\end{center}
\caption{Effective capacity in successive AF relays with the weak and short-term IRI constraints with $\bar{\gamma}=20\textrm{dB}$ and $q_0=20\textrm{dB}$, $q_0=21\textrm{dB}$, $q_0=23\textrm{dB}$ and $q_0=25\textrm{dB}$.}
\label{fig:6}
\end{figure}

After that, the effective capacity with the weak but short-term IRI constraints (see eq. \eqref{eq:13a}) is plotted in Fig. \ref{fig:5} for $\bar{\gamma}=10\textrm{dB}$ and Fig. \ref{fig:6} for $\bar{\gamma}=20\textrm{dB}$ for different values of $q_0$. We can observe that the effective capacity increases when the amount of acceptable IRI is increased. Without strict constraint on the interference, the relay can transmit more power temporally to improve the total effective capacity or requested QoS. 

Now, we can compare the obtained results in Fig. \ref{fig:2} and \ref{fig:3} with the results of Fig. \ref{fig:5} and \ref{fig:6} accurately. Generally, the effective capacity with the long-term constraint outperforms the results with the short-term constraint. However, despite the obtained results, short-term constraint is necessary in some sensitive applications and can not be ignored ever. In addition, when $\theta$ increases and high QoS is required, the effective capacity of successive AF relay with short-term IRI constraint drops rapidly. Therefore, we can conclude that the performance with the high QoS and short-term constraint is not suitable at all, and long-term constraint for management of interference is recommended for this situation properly.

In a similar way, the effective capacity with the strong and short-term IRI constraints (see eq. \eqref{eq:13c}), can be drawn and compared with Fig. \ref{fig:4}. However to avoid duplication, this figure is not plotted here. 

\begin{figure}[h]
\begin{center}
\includegraphics[draft=false,width=\linewidth]{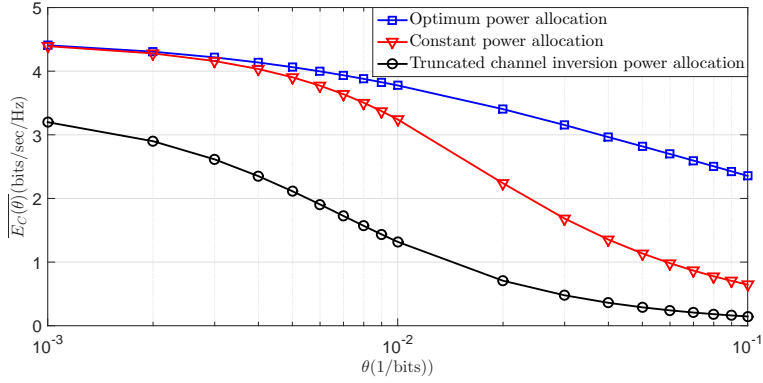}
\end{center}
\caption{Comparison of the effective capacity with different power allocation techniques when $\bar{\gamma}=20\textrm{dB}$ and $q_0=20\textrm{dB}$.}
\label{fig:7}
\end{figure}

For more details, in Fig. \ref{fig:7}, we have drawn the effective capacity of AF successive relays with the optimal power allocation \eqref{eq:16}, constant power allocation $\mu_0=1$ and truncated channel inversion with
\begin{equation}\label{eq:23}
\mu_0=\begin{cases}
0&,\gamma_{\mathrm{IR}}<\gamma_{\textrm{T}}\\
\frac{q_0}{\gamma_{\mathrm{IR}}}&,\gamma_{\mathrm{IR}}\geq \gamma_{\textrm{T}}
\end{cases}
\end{equation}
where $\gamma_{\textrm{T}}$ is obtained from $\textsf{E}\left\{\mu_0\right\}=1$. In constant power allocation technique, there is no constraint on the IRI. So, the performance at low value of $\theta$ with loose QoS, tends to the performance with the optimal power allocation. However, at high value of $\theta$, the performance with the optimal power allocation, outperforms the other techniques of transmission. In addition, when truncated channel inversion technique is used, the IRI can remain constant but the effective capacity performance is very poor. Therefore, the optimal power allocation (similar to \eqref{eq:16}) at the relay is strongly recommended specially when high QoS is required.

\begin{figure}[h]
\begin{center}
\includegraphics[draft=false,width=\linewidth]{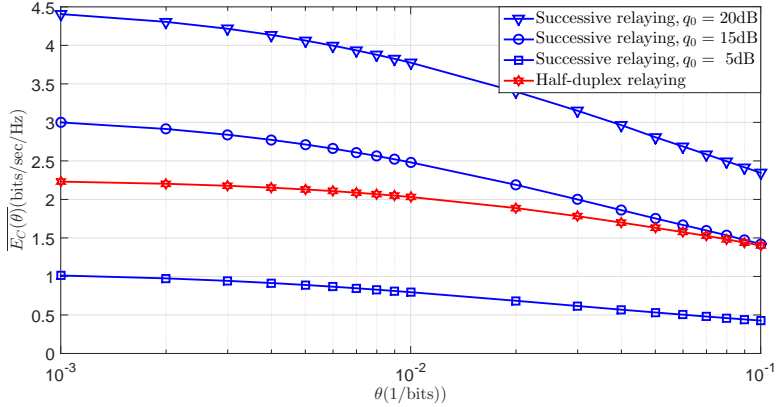}
\end{center}
\caption{Comparison of the effective capacity with the successive AF relays and HD relays when $\bar{\gamma}=20\textrm{dB}$.}
\label{fig:8}
\end{figure}

Finally, the effective capacity of successive AF relying scheme is compared with the traditional HD  relaying in Fig. \ref{fig:8} versus QoS exponent $\theta$. For the successive relays we assume the weak and long-term constraint and optimal power allocation \eqref{eq:16} with $q_0=5\textrm{dB}$ or $q_0=15\textrm{dB}$ or $q_0=20\textrm{dB}$. On the other hand, in the HD mode, there is no IRI and therefore, the optimal power allocation coefficient $\mu_0$ can be obtained from \eqref{eq:16} when $q_0$ tends to very small value (i.e. $q_0\to -\infty$). As we expected, we can see that successive relaying outperforms the traditional HD relaying especially when we manage the IRI with high value of threshold $q_0$ (i.e. $q_0=20\textrm{dB}$). When the threshold $q_0$ decreases and more strict constraint on the IRI is required, the effective capacity of successive relays decreases and approaches the effective capacity of HD relaying. However at low value of $q_0$ when the IRI is not tolerable, using HD relaying is more efficient than the successive relays.
\section{Conclusion}\label{sec:5}
In this paper, power allocation at the successive relay is studied. Successive relaying can be considered as a simple form of full-duplex relay implementation. In the successive relaying technique, two half-duplex relays with successive retransmission, act like a full-duplex relay. However, inter-relay interference management is crucial for proper operation of this technique. Here, the power allocation in AF successive relays for the effective capacity maximization was the goal. So, first, the power allocation coefficient with the long-term inter-relay interference constraint is calculated and a closed-form solution for the effective capacity is derived. Then, the power allocation with the short-term constraint is extracted. Finally, the effective capacity with the different type of power allocation is compared and we observed that in high QoS and low latency requirement, interference management with long-term constraint have a better performance and effective capacity results.
\appendix
\section{Proof of \eqref{eq:16}}\label{appx:1}
First, we can write the optimization problem in a standard form as
\begin{eqnarray}\label{a:1}
&\mathop{\max}_{\mu_0}-\textsf{E}\left\{\left(1+\mu_0 \gamma_{\mathrm{eq}}\right)^{-\tilde{\theta}}\right\}\nonumber\\
&\mathrm{s.t.}\quad\quad\textsf{E}\left\{\mu_0\right\}\leq 1,\quad\quad\textsf{E}\left\{\mu_0\gammaRR\right\}\leq q_0\nonumber\\
&-\mu_0\leq0.
\end{eqnarray}
Then, using Lagrangian method, the cost function is written as
\begin{equation}\label{a:2}
J=-\textsf{E}\left\{\left(1+\mu_0 \gamma_{\mathrm{eq}}\right)^{-\tilde{\theta}}\right\}+\lambda_1\left(1-\textsf{E}\left\{\mu_0\right\}\right)+\lambda_2\left(q_0-\textsf{E}\left\{\mu_0\gammaRR\right\}\right)+\lambda_3\mu_0
\end{equation}
where $\lambda_1\geq 0$, $\lambda_2\geq 0$ and $\lambda_3\geq 0$ are the nonnegative Lagrange multipliers corresponding to our constraints. Taking the partials with respect to $\mu_0$, we will have \cite{Opt}
\begin{eqnarray}\label{a:3}
\frac{\partial J}{\partial \mu}&=&\textsf{E}\left\{\tilde{\theta}\gamma_{\mathrm{eq}}\left(1+\mu_0 \gamma_{\mathrm{eq}}\right)^{-\tilde{\theta}-1}\right\}-\lambda_1-\lambda_2\textsf{E}\left\{\gammaRR\right\}+\lambda_3\nonumber\\
&=&\textsf{E}\left\{\tilde{\theta}\gamma_{\mathrm{eq}}\left(1+\mu_0 \gamma_{\mathrm{eq}}\right)^{-\tilde{\theta}-1}\right\}-\lambda_1-\lambda_2\bar{\gamma}+\lambda_3.
\end{eqnarray}
Now, we can find $\mu_0^{\mathrm{opt}}$, $\lambda_1^{\mathrm{opt}}$, $\lambda_2^{\mathrm{opt}}$ and $\lambda_3^{\mathrm{opt}}$ such that
\begin{equation}\label{a:4}
\textsf{E}\left\{\tilde{\theta}\gamma_{\mathrm{eq}}\left(1+\mu_0^{\mathrm{opt}} \gamma_{\mathrm{eq}}\right)^{-\tilde{\theta}-1}\right\}-\lambda_1^{\mathrm{opt}}-\lambda_2^{\mathrm{opt}}\bar{\gamma}+\lambda_3^{\mathrm{opt}}=0
\end{equation}
and
\begin{equation}\label{a:5}
\lambda_1^{\mathrm{opt}}\left(1-\textsf{E}\left\{\mu_0^{\mathrm{opt}}\right\}\right)=0
\end{equation}
\begin{equation}\label{a:6}
\lambda_2^{\mathrm{opt}}\left(q_0-\textsf{E}\left\{\mu_0^{\mathrm{opt}}\gammaRR\right\}\right)=0
\end{equation}
and 
\begin{equation}\label{a:7}
\lambda_3^{\mathrm{opt}}\mu_0^{\mathrm{opt}}=0
\end{equation}
correspondingly. From \eqref{a:7}, since $\mu_0^{\mathrm{opt}}=0$ is not acceptable, we can conclude that $\lambda_3^{\mathrm{opt}}=0$. Consider \eqref{a:5} and \eqref{a:6}, we can break the analysis into four different cases as follows.
\begin{enumerate}
\item{If $\lambda_1^{\mathrm{opt}}=0 \left(1-\textsf{E}\left\{\mu_0^{\mathrm{opt}}\right\}\neq0\right)$ and $\lambda_2^{\mathrm{opt}}=0 \left(q_0-\textsf{E}\left\{\mu_0^{\mathrm{opt}}\gammaRR\right\}\neq 0\right)$, then \eqref{a:4} converts to $\textsf{E}\left\{\tilde{\theta}\gamma_{\mathrm{eq}}\left(1+\mu_0^{\mathrm{opt}} \gamma_{\mathrm{eq}}\right)^{-\tilde{\theta}-1}\right\}=0$. Since we assume $\tilde{\theta}>0$, $\mu_0^{\mathrm{opt}}>0$, $\gamma_{\mathrm{eq}}>0$ and $\gammaRR>0$, this case is not a feasible solution.}
\item{If $\lambda_1^{\mathrm{opt}}=0\left(1-\textsf{E}\left\{\mu_0^{\mathrm{opt}}\right\}\neq0\right)$ and $q_0-\textsf{E}\left\{\mu_0^{\mathrm{opt}}\gammaRR\right\}=0 (\lambda_2^{\mathrm{opt}}\neq0)$, then \eqref{a:4} converts to $\textsf{E}\left\{\tilde{\theta}\gamma_{\mathrm{eq}}\left(1+\mu_0^{\mathrm{opt}} \gamma_{\mathrm{eq}}\right)^{-\tilde{\theta}-1}\right\}=\lambda_2^{\mathrm{opt}}\bar{\gamma}$. Here, we can assume $\tilde{\theta}\gamma_{\mathrm{eq}}\left(1+\mu_0^{\mathrm{opt}} \gamma_{\mathrm{eq}}\right)^{-\tilde{\theta}-1}=\lambda_2^{\mathrm{opt}}\bar{\gamma}$ and $\mu_0^{\mathrm{opt}}>0$ which leads to
\begin{equation}\label{a:8}
\mu_0^{\mathrm{opt}}=
\begin{cases}
0&,\gamma_{\mathrm{eq}}<\frac{\lambda_2^{\mathrm{opt}}\bar{\gamma}}{\tilde{\theta}}\\
\left(\frac{\tilde{\theta}}{\lambda_2^{\mathrm{opt}}\bar{\gamma}}\right)^\frac{1}{\tilde{\theta}+1}\left(\frac{1}{\gamma_{\mathrm{eq}}}\right)^\frac{\tilde{\theta}}{\tilde{\theta}+1}-\frac{1}{\gamma_{\mathrm{eq}}}&,\gamma_{\mathrm{eq}}\geq \frac{\lambda_2^{\mathrm{opt}}\bar{\gamma}}{\tilde{\theta}}
\end{cases}.
\end{equation}
Note that, $\lambda_2^{\mathrm{opt}}$ can be calculated from $\textsf{E}\left\{\mu_0^{\mathrm{opt}}\gammaRR\right\}=\textsf{E}\left\{\mu_0^{\mathrm{opt}}\right\}\bar{\gamma}=q_0$ or equivalently $\textsf{E}\left\{\mu_0^{\mathrm{opt}}\right\}=q_0/\bar{\gamma}$. Since we assumed $\lambda_1^{\mathrm{opt}}=0$, then we have $\textsf{E}\left\{\mu_0^{\mathrm{opt}}\right\}<1$. Therefore, this case is valid when $q_0/\bar{\gamma}<1$.
}
\item{If $\lambda_2^{\mathrm{opt}}=0\left(q_0-\textsf{E}\left\{\mu_0^{\mathrm{opt}}\gammaRR\right\}\neq 0\right)$ and $1-\textsf{E}\left\{\mu_0^{\mathrm{opt}}\right\}=0(\lambda_1^{\mathrm{opt}}\neq0)$, then \eqref{a:4} converts to $\textsf{E}\left\{\tilde{\theta}\gamma_{\mathrm{eq}}\left(1+\mu_0^{\mathrm{opt}} \gamma_{\mathrm{eq}}\right)^{-\tilde{\theta}-1}\right\}=\lambda_1^{\mathrm{opt}}$. Here, we can assume $\tilde{\theta}\gamma_{\mathrm{eq}}\left(1+\mu_0^{\mathrm{opt}} \gamma_{\mathrm{eq}}\right)^{-\tilde{\theta}-1}=\lambda_1^{\mathrm{opt}}$ and $\mu_0^{\mathrm{opt}}>0$ which leads to
\begin{equation}\label{a:9}
\mu_0^{\mathrm{opt}}=
\begin{cases}
0&,\gamma_{\mathrm{eq}}<\frac{\lambda_1^{\mathrm{opt}}}{\tilde{\theta}}\\
\left(\frac{\tilde{\theta}}{\lambda_1^{\mathrm{opt}}}\right)^\frac{1}{\tilde{\theta}+1}\left(\frac{1}{\gamma_{\mathrm{eq}}}\right)^\frac{\tilde{\theta}}{\tilde{\theta}+1}-\frac{1}{\gamma_{\mathrm{eq}}}&,\gamma_{\mathrm{eq}}\geq \frac{\lambda_1^{\mathrm{opt}}}{\tilde{\theta}}
\end{cases}.
\end{equation}
Note that, $\lambda_1^{\mathrm{opt}}$ can be calculated from $\textsf{E}\left\{\mu_0^{\mathrm{opt}}\right\}=1$. Since we assumed $\lambda_2^{\mathrm{opt}}=0$, then we have $\textsf{E}\left\{\mu_0^{\mathrm{opt}}\gammaRR\right\}=\textsf{E}\left\{\mu_0^{\mathrm{opt}}\right\}\bar{\gamma}<q_0$. Therefore, this case is valid when $q_0/\bar{\gamma}>1$.
}
\item{If $1-\textsf{E}\left\{\mu_0^{\mathrm{opt}}\right\}=0(\lambda_1^{\mathrm{opt}}\neq0)$ and $q_0-\textsf{E}\left\{\mu_0^{\mathrm{opt}}\gammaRR\right\}=0(\lambda_2^{\mathrm{opt}}\neq0)$, then \eqref{a:4} converts to $\textsf{E}\left\{\tilde{\theta}\gamma_{\mathrm{eq}}\left(1+\mu_0^{\mathrm{opt}} \gamma_{\mathrm{eq}}\right)^{-\tilde{\theta}-1}\right\}=\lambda_1^{\mathrm{opt}}+\lambda_2^{\mathrm{opt}}\bar{\gamma}$. Here, we can assume $\tilde{\theta}\gamma_{\mathrm{eq}}\left(1+\mu_0^{\mathrm{opt}} \gamma_{\mathrm{eq}}\right)^{-\tilde{\theta}-1}=\lambda_1^{\mathrm{opt}}+\lambda_2^{\mathrm{opt}}\bar{\gamma}=\ell$ and $\mu_0^{\mathrm{opt}}>0$ which leads to
\begin{equation}\label{a:10}
\mu_0^{\mathrm{opt}}=
\begin{cases}
0&,\gamma_{\mathrm{eq}}<\frac{\ell}{\tilde{\theta}}\\
\left(\frac{\tilde{\theta}}{\ell}\right)^\frac{1}{\tilde{\theta}+1}\left(\frac{1}{\gamma_{\mathrm{eq}}}\right)^\frac{\tilde{\theta}}{\tilde{\theta}+1}-\frac{1}{\gamma_{\mathrm{eq}}}&,\gamma_{\mathrm{eq}}\geq \frac{\ell}{\tilde{\theta}}
\end{cases}.
\end{equation}
Note that, $\ell$ can be calculated from both $\textsf{E}\left\{\mu_0^{\mathrm{opt}}\right\}=1$ or $\textsf{E}\left\{\mu_0^{\mathrm{opt}}\gammaRR\right\}=\textsf{E}\left\{\mu_0^{\mathrm{opt}}\right\}\bar{\gamma}=q_0$ with the same value. Therefore, this case is valid when $q_0/\bar{\gamma}=1$.
}
\end{enumerate}
Now, we can aggregate cases 2, 3 and 4 in one case such as
\begin{equation}\label{a:11}
\mu_0^{\mathrm{opt}}=
\begin{cases}
0&,\gamma_{\mathrm{eq}}<\gamma_{\textrm{T}}\\
\left(\frac{1}{\gamma_{\textrm{T}}}\right)^\frac{1}{\tilde{\theta}+1}\left(\frac{1}{\gamma_{\mathrm{eq}}}\right)^\frac{\tilde{\theta}}{\tilde{\theta}+1}-\frac{1}{\gamma_{\mathrm{eq}}}&,\gamma_{\mathrm{eq}}\geq \gamma_{\textrm{T}}
\end{cases}.
\end{equation}
where $\gamma_{\textrm{T}}$ is a cut-off SNR threshold which is determined from 
\begin{equation}\label{a:12}
\textsf{E}\left\{\mu_0^{\mathrm{opt}}\right\}=
\begin{cases}
\frac{q_0}{\bar{\gamma}}&,q_0<\bar{\gamma}\\1&,q_0\geq\bar{\gamma}
\end{cases}.
\end{equation}
\section{Proof of \eqref{eq:19}}\label{appx:2}
By substituting \eqref{eq:16} or \eqref{eq:18} into \eqref{eq:15}, we will have
\begin{equation}\label{a:13}
\textsf{E}\left\{\left(1+\mu_0^{\mathrm{opt}}\gamma_{\mathrm{eq}}\right)^{-\tilde{\theta}}\right\}=\underbrace{\int_{0}^{\gamma_{\textrm{T}}}f_{\gamma_{\mathrm{eq}}}(x)dx}_{I_1}+\underbrace{\int_{\gamma_{\textrm{T}}}^{\infty}\left(x/\gamma_{\textrm{T}}\right)^{-\tilde{\theta}/(1+\tilde{\theta})}f_{\gamma_{\mathrm{eq}}}(x)dx}_{I_2}
\end{equation}
where $f_{\gamma_{\mathrm{eq}}}(x)$ is presented in \eqref{eq:2}. For solving $I_1$,  it can be changed to two infinite integrals as
\begin{eqnarray}\label{a:14}
I_1&=&\underbrace{\int_{0}^{\infty}f_{\gamma_{\mathrm{eq}}}(x)dx}_{I_3}-\underbrace{\int_{\gamma_{\textrm{T}}}^{\infty}f_{\gamma_{\mathrm{eq}}}(x)dx}_{I_4}\nonumber\\
&=&\frac{\sqrt{\pi}}{2\Gamma(5/2)}\left[F(3,3/2,5/2;0)+\frac{1}{2}F(2,1/2,5/2;0)\right]\nonumber\\
&-&\frac{\sqrt{\pi}\gamma_{\textrm{T}}}{\bar{\gamma}}\left[G_{23}^{30}\left(\frac{4\gamma_{\textrm{T}}}{\bar{\gamma}}\left|\begin{array}{l}0,3/2\\-1,2,0\end{array} \right.\right) + G_{23}^{30}\left(\frac{4\gamma_{\textrm{T}}}{\bar{\gamma}}\left|\begin{array}{l}0,3/2\\-1,1,1\end{array} \right.\right)\right]
\end{eqnarray}
where the closed-form solution of $I_3$ and $I_4$ is possible using \cite[eq. 6.621-3]{Ryzhik} and \cite[eq. 6.625-7]{Ryzhik}, respectively. Once again \cite[eq. 6.625-7]{Ryzhik} can be used for finding $I_2$. Therefore, we have
\begin{eqnarray}\label{a:15}
I_2&=&\frac{\sqrt{\pi}}{4}\left(\frac{4\gamma_{\textrm{T}}}{\bar{\gamma}}\right)^{\frac{1+2\tilde{\theta}}{1+\tilde{\theta}}}\left[G_{23}^{30}\left(\frac{4\gamma_{\textrm{T}}}{\bar{\gamma}}\left|\begin{array}{l}0,\frac{1}{2}+\frac{1}{1+\tilde{\theta}}\\-1,1+\frac{1}{1+\tilde{\theta}},-1+\frac{1}{1+\tilde{\theta}}\end{array} \right.\right)\right.\nonumber\\
&+& \left.G_{23}^{30}\left(\frac{4\gamma_{\textrm{T}}}{\bar{\gamma}}\left|\begin{array}{l}0,\frac{1}{2}+\frac{1}{1+\tilde{\theta}}\\-1,\frac{1}{1+\tilde{\theta}},\frac{1}{1+\tilde{\theta}}\end{array} \right.\right)\right]
\end{eqnarray}
where $F(.,.;.;.)$ represents Gauss hypergeometric function \cite[eq. 9.10]{Ryzhik}, $\Gamma(.)$ denotes the Gamma function and $G_{p,q}^{m,n}(.)$ is the Meijer's G function defined in \cite[eq. 9.301]{Ryzhik}. Now, connecting the obtained results for $I_1$ and $I_2$, we have
\begin{eqnarray}\label{a:16}
E_C(\theta)&=&-\frac{1}{\theta}\ln\left\{\frac{\sqrt{\pi}}{2\Gamma(5/2)}\left[F(3,3/2,5/2;0)+\frac{1}{2}F(2,1/2,5/2;0)\right]\right.\nonumber\\
&-&\frac{\sqrt{\pi}\gamma_{\textrm{T}}}{\bar{\gamma}}\left[G_{23}^{30}\left(\frac{4\gamma_{\textrm{T}}}{\bar{\gamma}}\left|\begin{array}{l}0,3/2\\-1,2,0\end{array} \right.\right) + G_{23}^{30}\left(\frac{4\gamma_{\textrm{T}}}{\bar{\gamma}}\left|\begin{array}{l}0,3/2\\-1,1,1\end{array} \right.\right)\right]\nonumber\\
&+&\left.\frac{\sqrt{\pi}}{4}\left(\frac{4\gamma_{\textrm{T}}}{\bar{\gamma}}\right)^{\frac{1+2\tilde{\theta}}{1+\tilde{\theta}}}\left[G_{23}^{30}\left(\frac{4\gamma_{\textrm{T}}}{\bar{\gamma}}\left|\begin{array}{l}0,\frac{1}{2}+\frac{1}{1+\tilde{\theta}}\\-1,1+\frac{1}{1+\tilde{\theta}},-1+\frac{1}{1+\tilde{\theta}}\end{array} \right.\right)\right.\right.\nonumber\\
&+&\left.\left.G_{23}^{30}\left(\frac{4\gamma_{\textrm{T}}}{\bar{\gamma}}\left|\begin{array}{l}0,\frac{1}{2}+\frac{1}{1+\tilde{\theta}}\\-1,\frac{1}{1+\tilde{\theta}},\frac{1}{1+\tilde{\theta}}\end{array} \right.\right)\right]\right\}.
\end{eqnarray}


\end{document}